%% file: main.tex
\documentclass[twoside,twocolumn]{article}

\usepackage{textcomp}

\usepackage{xspace}

\usepackage{hyperref}
\usepackage{gensymb}
\usepackage{float}
\usepackage{multicol}
\usepackage{physics}
\usepackage{siunitx}
\usepackage{amsfonts}
\usepackage{amssymb}
\usepackage{verbatim}
\usepackage{hyperref}
\usepackage{url} 
\usepackage{graphicx}
\usepackage{mathptmx}
\usepackage{amsmath}
\usepackage{mathtools}
\usepackage{cleveref}
\crefname{equation}{Eq.}{Eqs.}
\Crefname{equation}{Equation}{Equations}
\creflabelformat{equation}{#2#1#3}
\crefrangeformat{equation}{#3#1#4--#5#2#6}
\usepackage[numbers,sort&compress]{natbib}
\DeclareMathAlphabet{\mathcal}{OMS}{cmsy}{m}{n} 
\usepackage{blindtext} 

\usepackage[english]{babel} 

\usepackage[hmarginratio=1:1, left = 18mm, top=32mm,columnsep=20pt]{geometry} 
\usepackage[hang, small,labelfont=bf,up,textfont=it,up]{caption} 
\usepackage{booktabs} 

\usepackage{abstract} 

\usepackage{titlesec} 
\renewcommand\thesection{\Roman{section}} 
\renewcommand\thesubsection{\roman{subsection}} 
\titleformat{\section}[block]{\large\scshape\centering}{\thesection.}{1em}{} 
\titleformat{\subsection}[block]{\large}{\thesubsection.}{1em}{} 

\usepackage{titling} 

\usepackage{hyperref} 

\usepackage{soul}   

\title{Visualization of time-frequency structures in gravitational wave signals} 
\author{%
\textsc{Chad Henshaw, Megan Arogeti, Alice Heranval, Laura Cadonati} \\[1ex] 
\normalsize \emph{School of Physics, Georgia Institute of Technology, Atlanta, GA 30332, USA}
\normalsize 
}
\date{\today} 
\renewcommand{\maketitlehookd}{%
\begin{abstract}
\noindent The gravitational wave signals produced by the coalescence of compact binaries progress through three stages: inspiral, merger, and postmerger. The evolution of their frequency follows a slow build up during the inspiral that peaks at merger, forming the characteristic ``chirp" pattern in the signal's time-frequency map. Herein we introduce a framework for localizing further characteristic structures in the time-frequency space of gravitational wave signals using the continuous wavelet transform. We consider two example cases where there are specific patterns in the postmerger stage of the signal that are rich with information on the physical nature of the source: highly-inclined black hole binaries with asymmetric mass ratio, and neutron star binaries with postmerger remnant oscillations. It is demonstrated that the choice of quality factor $Q$ plays a central role in distinguishing the postmerger features from that of the inspiral, with black hole systems preferring lower $Q$ and neutron star systems preferring higher $Q$. Furthermore, we demonstrate the use of chirplets as the wavelet transform basis, which allow for manipulation of structure in the time-frequency map.

\end{abstract}
}

\begin{document}

\maketitle

\section{Introduction}

Gravitational waves (GWs) from compact binary coalescences (CBCs) exhibit distinctive characteristics in the evolution of their frequency. To date, the LIGO, Virgo, and KAGRA scientific collaborations~\cite{Aasi2015, Acernese2015, Akutsu2021}, which now comprise the LIGO-Virgo-KAGRA Collaboration (LVK), have detected 90+ signals from CBC systems~\cite{gwtc1, gwtc2, gwtc2.1, gwtc3}. Starting at some initial separation distance, the two objects in the binary orbit a common center of mass, and as time progresses their separation distance will decrease as energy is lost through gravitational radiation in the form of GWs. As the separation distance decreases, the orbital period decreases, thus the orbital frequency increases, and thus the GW frequency also increases $\left(f_{gw} \approx 2 f_{orb.}\right)$. During this stage of the merger event, known as the inspiral, the GW amplitude also increases, peaking at the time of merger, and therefore the frequency evolution during the inspiral is referred to as a ``chirp." At lowest post-Newtonian (PN) order, the GW frequency during the inspiral evolves as $\dot{f}(t) \propto \mathcal{M}^{5/3} f^{11/3}$~\cite{Cutler1994} where $\mathcal{M} = M \left[\frac{q}{\left(1+q
\right)^2}\right]^{3/5}$ is the ``chirp mass," $M = m_1 + m_2$ is the total mass, and $q = m_1 /m_2, \: m_1 \geq m_2$ is the mass ratio. The inclusion of higher order terms also sees contribution from the mass ratio, the spin angular momenta of the compact objects, and their tidal deformabilities in the case of objects with matter like neutron stars (NSs). Near the point of merger this approximation breaks down, and a fully relativistic evaluation is needed to describe the merger. The characterization of the postmerger stage varies depending on the nature of the compact objects in the binary. If one of the two objects is a black hole (BH), then the merger constitutes the formation of a distorted common horizon that encapsulates both objects, which then ``rings down" in the postmerger to spherical symmetry.\par 

The frequency evolution of a GW signal can reveal much about the nature of the originating system, as GWs fundamentally encode the properties of their source. Time-frequency analysis sees wide use in the gravitational wave community, over a variety of applications. One of the primary applications for time-frequency mapping in gravitational wave data analysis is the unmodeled search for transient signals. This effort does not assume anything about the shape of the gravitational wave signal, and instead relies on time-frequency analysis utilizing the multi-resolution ``Q-scan," introduced in~\cite{Chatterji2004, Chatterji2005}. If the detector noise is assumed to be stationary and Gaussian, then the presence of any other component - like a gravitational wave signal, or other noise transient - will show up in the time-frequency map as excess power \cite{Anderson2001}. This excess power method has been widely used in all-sky searches of gravitational wave data \cite{allsky_LIGO-GEO-Virgo, allsky_O1, allsky_O2, allsky_O3}, which incorporate both online \cite{Klimenko2016} and offline follow-up \cite{Robinet2020, Cornish2021} algorithms. Time-frequency analysis is also commonly used in the identification of noise transients in LVK data for detector characterization~\cite{Robinet2020, Zevin2017}.\par

Beyond these applications, time-frequency analysis can also be used to understand specific frequency characteristics in gravitational waves that are directly connected to the physics of their originating system. In this document we will review two examples of systems that contain distinctive frequency structure in the post-merger gravitational wave signal: binary neutron star (BNS) systems, and highly-inclined binary black hole (BBH) systems with asymmetric mass ratio. By applying targeted time-frequency analysis using the continuous wavelet transform (CWT) to synthetic representations of these signals, we will demonstrate that these features may be isolated from the primary frequency spike due to the inspiral. This enables routes to further analysis such as parameterization for theoretical exploration~\cite{Henshaw2025}, or detection using stacking methods which will be discussed in a subsequent publication.\par


In a recent Numerical Relativity (NR) study~\cite{CalderonBustillo2020}, it was demonstrated that GW signals from BBH systems with asymmetric mass ratio $q > 1$ exhibit additional frequency peaks after merger when viewed from a highly-inclined angle. This study further demonstrates that these frequency peaks are correlated to regions of tight curvature on the horizon of the distorted final BH. Such signals are therefore of immediate interest, as if this correlation can be understood then there is a possibility of inferring the final black hole horizon geometry from the gravitational wave signals that we receive on Earth. We direct the reader to our companion work~\cite{Henshaw2025} for more details. Additionally, as these postmerger frequency peaks can be easily mistaken for spurious transients (glitches) in detector data~\cite{Sharma2022}, understanding how to differentiate them as a physical effect is important for the realistic detection of intermediate-mass black hole (IMBH) binary systems.\par  



In the case of a BNS system, the postmerger gravitational wave signal is highly variable depending on several factors. Neutron stars are highly dense, compact stellar objects where the interior structure is dictated by an equation of state (EoS), which provides a relationship between the NS interior pressure, density, and temperature. The exact nature of these conditions in the interior of neutron stars is unknown\cite{EOSDEP, Torres-Rivas:2018svp}, but gravitational waves signals from BNS mergers offer a means by which the EoS may be studied. The EoS is encoded not only in the inspiral stage of a BNS merger \cite{Chatziioannou:2018vzf, GW170817EoS}, but also in the postmerger stage. Depending on the component masses and the EoS, a BNS merger could result in i) a prompt collapse to a black hole, ii) a hypermassive or supramassive neutron star that undergoes a delayed collapse to a black hole or iii) a stable neutron star \cite{Clark2016, Postmerger2, Chatziioannou:2017ixj}. In general, oscillations of this postmerger remnant can emit a high-frequency gravitational wave signal, the specific characteristics of which are dependent on the remnant properties. In the case of a delayed collapse to a black hole, the post-merger gravitational wave signal could have a duration on the order of  $10-100$ [ms], with frequencies between $1 - 4$ [kHz]. If detected, this postmerger signal could provide a window to study a higher-mass regime than allowed by the inspiral signal \cite{Postmerger2, Chatziioannou:2017ixj}.\par

This document is organized as follows. In Sec.(\ref{methods}), we will briefly review the fundamentals of time-frequency analysis, and in Sec.(\ref{CWT_section}) we introduce an implementation of the CWT method. In Sec.(\ref{applications}), we will then apply the CWT to select simulated BBH and BNS postmerger waveforms to demonstrate its use for feature discrimination. Finally, we conclude in Sec.(\ref{discussion}) with a discussion on further applications and routes of analysis.

\section{Precepts of time-frequency analysis}
\label{methods}

The Fourier transform takes a continuous time series $x(t)$ and decomposes it into its constituent frequencies, rendering a continuous frequency series $\hat{x}(f)$:

\begin{align}
    \hat{x}(f) = \int_{-\infty}^{\infty} x(t) \exp\left[-i 2\pi f t\right] dt.
    \label{FT}
\end{align}

Considering the product between the time series $x(t)$ and the complex sinusoid $\exp\left[-i 2\pi f t\right]$ in Eq.(\ref{FT}), the Fourier transform is a check for similarity. The sinusoid can be thought of as an \emph{analyzing function} $g(t)$, and thus the integral in Eq.(\ref{FT}) is the inner product:

\begin{align}
    \langle x, g \rangle = \int_{-\infty}^{\infty} x(t) g^*(t) dt,
\end{align}

where $g^*(t)$ denotes the complex conjugate of $g(t)$\footnote{Note that in Eq.(\ref{FT}) the complex conjugate of the exponential has already been taken, hence the negative exponent.}. The resultant complex function $\hat{x}(f)$ describes the amplitude and phase of the component of $x(t)$ that matches the frequency $f$ of the sinusoid, which spans a continuous domain. This frequency domain representation $\hat{x}(f)$ can likewise be converted back to the time domain by the inverse Fourier transform:

\begin{align}
    x(t) = \int_{-\infty}^{\infty} \hat{x}(f) \exp\left[i 2\pi f t\right] df,
\end{align}

such that $x(t)$ and $\hat{x}(f)$ form a Fourier pair. Note that in performing the conversion from $x(t)$ to $\hat{x}(f)$, all information about the signal's time evolution is forsaken. Therefore given a signal $x(t)$, the Fourier transform can describe the signal's \emph{spectrum} (the strength of its frequency components), but it does not describe \emph{when} in time the signal reaches those frequencies. To understand simultaneously the signal's frequency evolution, and its strength per frequency, one needs to compute a two-dimensional function $X(t, f)$ that preserves both the time and frequency information. One of the primary methods for time-frequency analysis is the short-time Fourier transform, first devised by Dennis Gabor in 1946~\cite{Gabor1946}:

\begin{align}
    X(\tau, f) = \int_{-\infty}^{\infty} x(t) w^*(t - \tau) \exp\left[- i 2\pi f t\right] dt,
    \label{STFT}
\end{align}

where $w(\tau)$ is a window function centered on $\tau = 0$; i.e. a function that is only non-zero within some chosen interval. Note that Eq.(\ref{STFT}) is exactly a FT as in Eq.(\ref{FT}), where the effective time series $x(t)$ is now the product between the signal and the time-shifted window $w(t - \tau)$. One can also think of the inner product picture, where the analyzing function is now the product between the window and the sinusoid. Under either interpretation, the introduction of the time-shift changes the Fourier transform into a \emph{convolution}, in which the analyzing function ``slides" along the signal, and the overlap $X(\tau, f)$ is computed at each (continuous) time $\tau$ and (continuous) frequency $f$. Because the window has a finite width, this is effectively computing the FT of the signal over a short time segment, then stacking those segments next to each other to build a continuous two-dimensional time-frequency representation column by column. Alternatively in the inner product picture, the convolution produces a new time series of coefficients at the central frequency of the analyzing function, essentially comprising one row in the resultant two-dimensional time-frequency map.\par

There are several different types of window functions that may be used with the STFT, but one of the drawbacks of this method is that the window function has a fixed width. Regardless of the specific function, the window width is related to the time-resolution of the resultant spectrogram. The larger the window duration, the poorer one's time resolution, but the finer the frequency resolution becomes. Because the width of the window is fixed, the STFT is limited in its ability to adapt to signals with rapid changes in frequency evolution. In contrast, the CWT method discussed below allows for variable window widths, and therefore has greater flexibility in adapting to the signal.


\section{The continuous wavelet transform}
\label{CWT_section}

The continuous wavelet transform (CWT) is a process that scans a type of filter called a \emph{wavelet} - an oscillating function modulated by some envelope - across a signal to probe its time and frequency space. For more detail on wavelets see Appendix Sec.(\ref{appendix_wavelets}). The CWT is defined by the following expression:

\begin{align}
    T(a,b) = \int_{-\infty}^{\infty} x(t) \left[w(a)\psi^{*}\left(\frac{t - b}{a}\right)\right] dt,
    \label{CWT}
\end{align}

where $x(t)$ represents the signal, $w(a)$ is a weighting function\footnote{$w(a)$ is usually set such that wavelets at any $a$ will have the same energy.}, and $\psi^{*}$ is the complex conjugate of the `mother wavelet' - the wavelet being scanned across the signal. The scanning occurs by scaling the wavelet by a dilation parameter $a$, and shifting its time localization by a translation parameter $b$. As with the STFT, the time shifting makes Eq.(\ref{CWT}) a convolution between the scaled wavelet function - i.e. the analyzing function - and the signal. The CWT coefficients $T(a,b)$ may then be computed in the same manner as previously, but now instead of passing explicit frequencies to define our filter we pass an array of discrete wavelet scales. In practice the signal data is a digital time series with sampling rate $f_s$ and time steps $\Delta b = f_s^{-1}$. For each scale $a$, we fast Fourier transform (FFT) both the wavelet and the signal, take their product in Fourier space, then inverse FFT the result to produce a time series array of coefficients per $a$.\footnote{In this sense the continuity of the transform is approximated by discretization given by the spacing between wavelet scales and the sampling rate of the signal.} The resultant two-dimensional map is a \emph{scalogram} - as it maps the scaled wavelets to their inner product with the signal at central times $b$. The wavelet scales $a$ may then be related to their central frequencies, and in this way the CWT builds the time-frequency map row by row.\par

Alternatively, one may also imagine the CWT process as a point-wise inner product between the wavelet and the signal. For each time step $b$, the wavelet's scale is varied by the dilation parameter $a$. These parameters therefore become coordinates that map the signal's time-frequency representation. Under this interpretation, the CWT is often referred to as a ``mathematical microscope," a metaphor that may be further extended by considering the wavelet form chosen for the operation as a type of lens. The dilation parameter $a$ therefore describes the width of the lens' aperture, and the translation parameter $b$ describes its location within the signal domain. Regardless of the interpretation, $T(a,b)$ is a measure of the similarity between the wavelet and the signal at the point $(a,b)$. In this way the CWT can be tuned to identify specific substructures in the time-frequency space of a signal, by choosing a mother wavelet that resembles those features. Let's now consider the example of the scaled and translated Morlet-Gabor (MG) wavelet, given in the time and frequency domains by:

\begin{align}
    &\psi(t; f_0, a, b) =\nonumber\\
    &= \pi^{-\frac{1}{4}} a^{-1/2} \exp\left[-\frac{1}{2}\left(\frac{t-b}{a}\right)^2\right]\exp\left[i 2 \pi f_0 \left(\frac{t-b}{a}\right) \right]. \label{MGwavelet_timedomain}
\end{align}
\begin{align}
    &\tilde{\psi}(f; f_0, a, b) =\nonumber\\
    &=\sqrt{2} \pi^{\frac{1}{4}} a^{1/2} \exp\left[-i 2\pi f b\right]\exp\left[-\frac{1}{2}\left(2\pi f_0 - 2\pi a f\right)^2\right]. \label{MGwavelet_freqdomain}    
\end{align}

This wavelet has unit energy, central time $\tau = b$, central frequency $\phi = f_0 / a$, characteristic duration $\sigma_t = a / \sqrt{2}$, and bandwidth $\sigma_f = 1 / 2\sqrt{2} \pi a$; see Appendix Sec.(\ref{appendix_wavelets}) for derivations. Note that the duration-bandwidth product is $\sigma_t \sigma_f = 1 / 4\pi$, so MG wavelets are maximally compact in time-frequency space; i.e. they reach the minimum time-frequency area allowed by the Heisenberg-Gabor uncertainty principle. The quality factor $Q$ is then given by:

\begin{align}
    Q \equiv \frac{\phi}{\sigma_f} = 2\sqrt{2} \pi f_0, \label{MG_Q}
\end{align}

which we see does not depend on the wavelet scale; $Q$ depends only on the starting frequency of the sinusoid in the mother wavelet. The time resolution is proportional to the wavelet scale, and the frequency resolution is inversely proportional to the wavelet scale. Given the relationship between wavelet scale and central frequency, this means that passing a linearly spaced array of wavelet scales into the CWT will yield better time localization at higher frequencies, at the cost of frequency resolution. As a corollary, the CWT will yield better frequency resolution at lower frequencies, and the cost of time localization.\par

For this reason the CWT is well suited for analyzing gravitational waves produced by CBC events that have rapidly changing frequency structure in the post-merger. The frequency evolution of such signals is characterized by a slow build at low frequency during the inspiral which peaks at merger. However for specific cases the post-merger signal can contain more complex structure in frequency evolution, which occur at higher overall frequency. Such cases are naturally suited for analysis with the CWT, as the slowly evolving behavior of the inspiral is less sensitive to the lower time resolution inherent at lower frequency, but the rapid changes in the post-merger that occur at higher frequency may be resolved. In Sec.(\ref{applications}) below, we investigate two such cases: oscillation frequencies in the post-merger BNS signal, and frequency peaks in the post-merger of highly-inclined asymmetric mass ratio BBH signals.



\section{Applications}
\label{applications}

\begin{figure}
    \centering
    \includegraphics[width=3.5in]{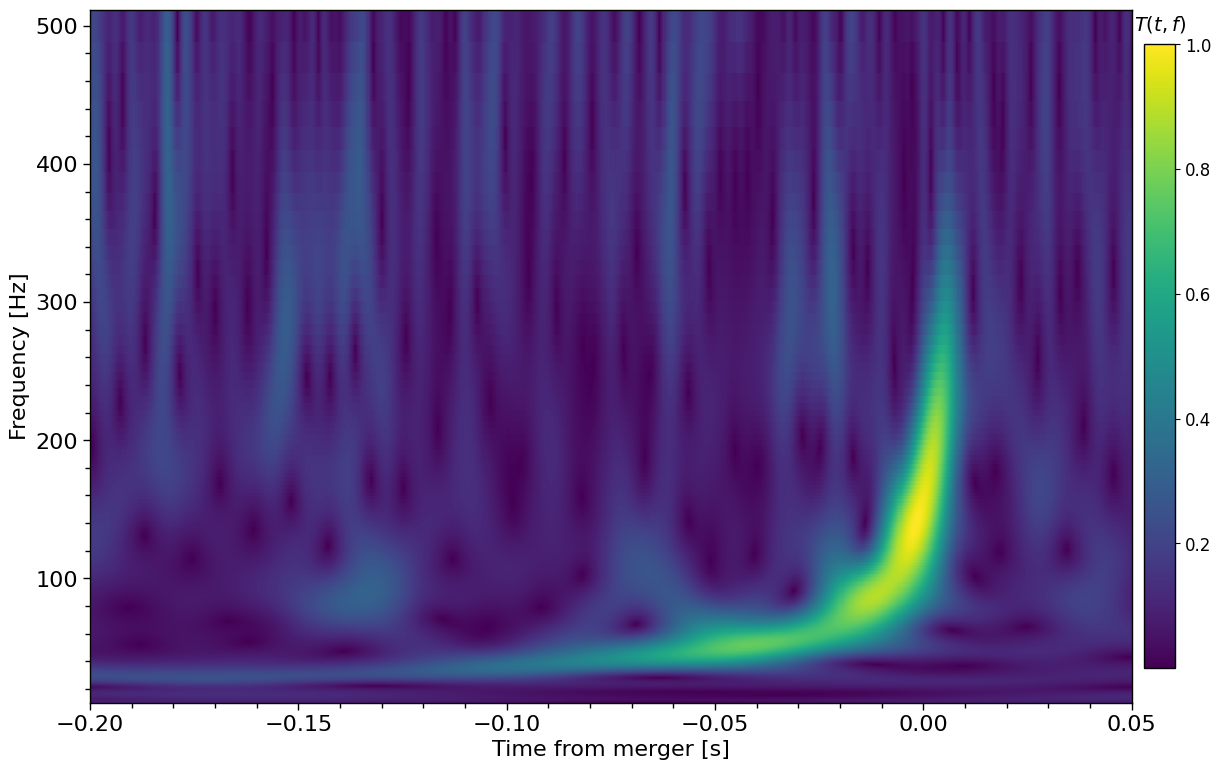}
    \caption{A time-frequency map of GW150914, the first detected gravitational wave signal, created using the CWT with Morlet-Gabor wavelets at $Q=6.0$ and logarithmic frequency spacing.}
    \label{GW150914_example}
\end{figure}

When applied to a gravitational wave signal, the CWT produces a time-frequency map, or scalogram, that captures the primary aspects of the signal's frequency evolution. Shown in Fig.(\ref{GW150914_example}) is an example scalogram for GW150914, the first gravitational wave signal detected by the LIGO and Virgo Collaborations~\cite{gw150914, gwtc1}. This image does not display the entire inspiral portion of the signal, but one can see that over the course of 0.2 seconds the frequency rapidly and smoothly evolves monotonically, increasing by a factor of $\sim 10$ before reaching the merger.\par 

To produce the image in Fig.(\ref{GW150914_example}), the data for GW150914 was obtained using the the CWT was run using Morlet-Gabor wavelets at $Q=6.0$, and the resultant scalogram was normalized by dividing all coefficient values by the largest value in the map. This makes the units in the color bar of Fig.(\ref{GW150914_example}) somewhat arbitrary, and as such this scaling method makes the analysis less useful for differentiating the signal from noise when compared to the Q-scan method that searches for excess coherent power~\cite{Chatterji2004, Chatterji2005, Anderson2001}. However, the consistent normalization facilitates exploration into how adjusting the parameters of the CWT analysis affects the structures in the time-frequency map for a freespace gravitational wave.   

\subsection{BBH Postmerger}
\label{BBH Postmerger}

Beyond the primary frequency peak denoting the inspiral-merger transition, gravitational waves from binary black hole systems with asymmetric mass ratio $q > 1$ exhibit additional frequency peaks after merger when viewed from a highly-inclined angle. These frequency structures were first studied in~\cite{CalderonBustillo2020}, where it was demonstrated that there is a correlation between the post-merger signal and the curvature of the final black hole horizon. Such signals are therefore of immediate interest, as if this correlation can be understood then there is a possibility of inferring the final black hole horizon geometry from the gravitational wave signals that we receive on Earth.\par

\begin{figure}
    \centering
    \includegraphics[width=3.5in]{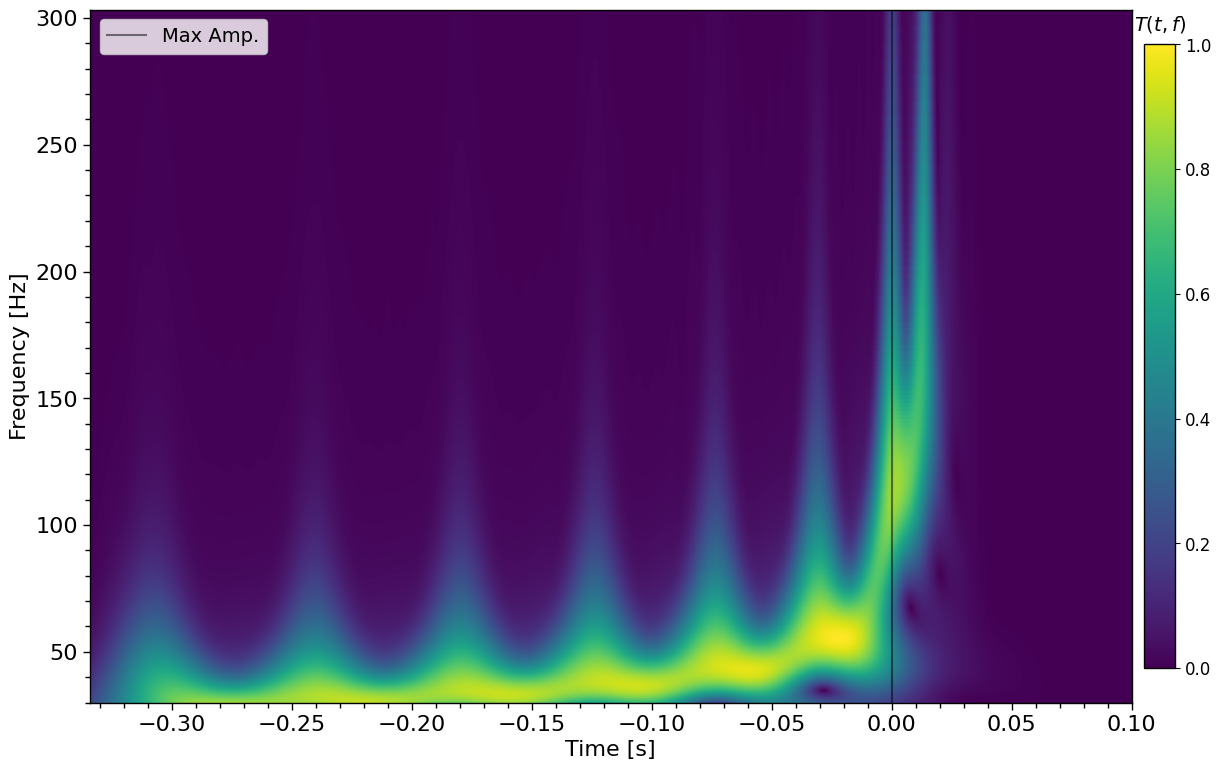}
    \caption{Frequency structure evolution in the noiseless, freespace gravitational waveform GT0568 $\left(q = 10.0,\: \abs{\mathbf{S}_1} = \abs{\mathbf{S}_2} = 0\right)$, scaled to $M = 80 \: M_{\odot}$ at $d_L = 500$ [Mpc], oriented at inclination $\iota = \pi / 2$ and phase $\phi = 0$, and generated from a starting frequency of 30.0 [Hz]. The black vertical line denotes the time location of the maximum waveform amplitude, which is used as an approximation for the time of merger.}
    \label{GT0568_Q5}
\end{figure}

As a first step towards investigating this possibility, it is critical to understand how to map the time-frequency space of these waveforms. For this purpose, we use the CWT to probe frequency structure. Shown in Fig.(\ref{GT0568_Q5}) is the time-frequency map of the NR waveform GT0568~\cite{Jani2016, Ferguson2023}, which is a zero-spin quasicircular BBH merger with mass ratio $q = 10$, that includes spherical harmonic modes up to $l=8$. One can see that in contrast to the relatively simple structure in Fig.(\ref{GW150914_example}), which is consistent with an equal mass ratio $(q\approx 1)$ system, the mass asymmetry in GT0568 introduces a more complex frequency structure when viewed from the edge-on inclination of $\pi / 2$ at a phase angle of $\phi = 0$. The waveform is shifted in time to locate the peak waveform amplitude at $t=0.0$, which we use as an approximation for the time of merger. One can see that after the initial frequency spike from the inspiral-merger transition, there is an additional frequency spike in the postmerger. See Sec.(\ref{discussion}) and our companion work~\cite{Henshaw2025} for further discussion on the physical interpretation of this feature, which we refer to as the double-chirp pattern.\par 

\begin{figure}
    \centering
    \includegraphics[width=3.5in]{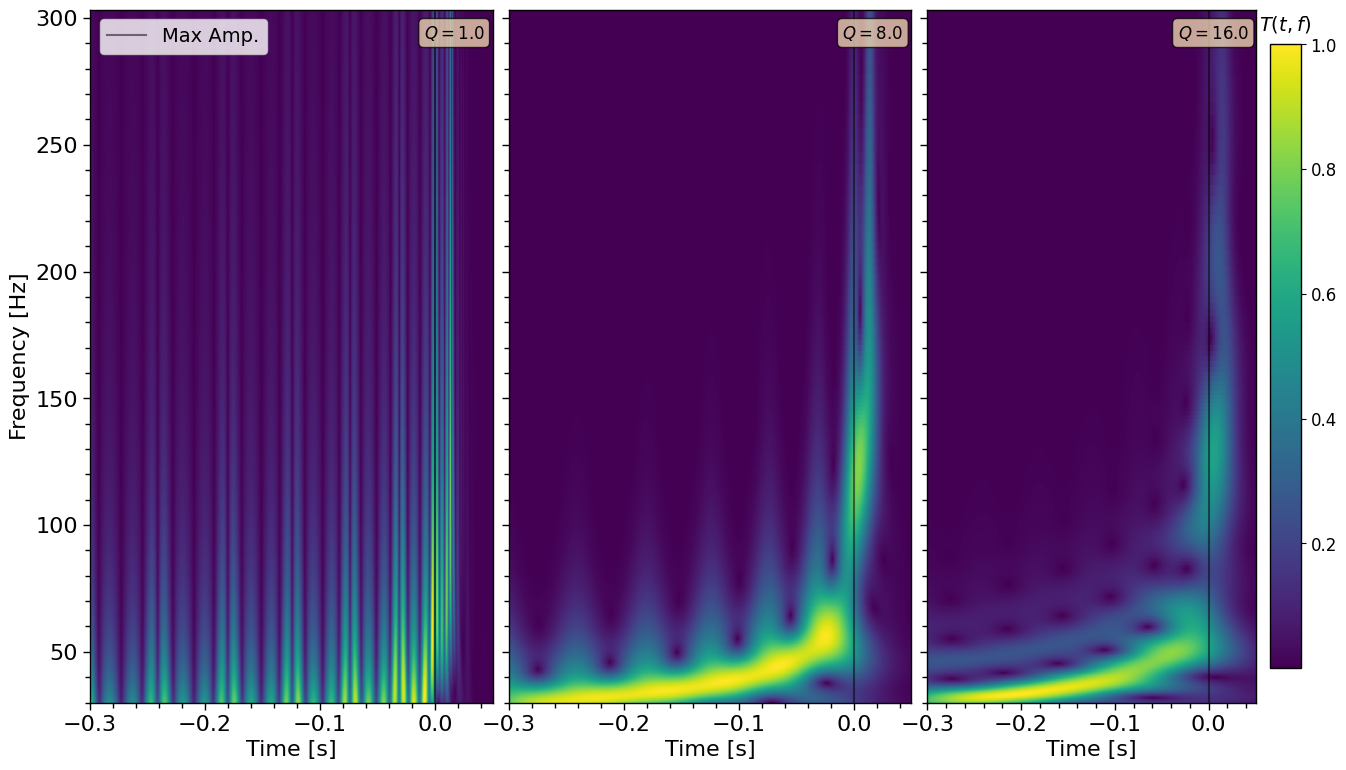}
    \caption{Quality factor variation for the CWT time-frequency map with the Morlet wavelet applied to the asymmetric mass ratio BBH waveform GT0568, viewed at edge-on inclination. From left to right, $Q = 1.0, 8.0, 16.0$.}
    \label{GT0568_morlet_Qvary}
\end{figure}

The scalogram in Fig.(\ref{GT0568_Q5}) was made using the CWT with the Morlet wavelet at $Q = 5.0$. As discussed in Sec.(\ref{methods}) above, the quality factor $Q$ not only defines the mother frequency for the wavelet, it is also related to the time and frequency resolutions of the CWT. For lower values of Q, the time resolution will be enhanced but the frequency resolution will be diminished, and vice versa. Towards the goal of identifying the postmerger frequency peak from that of the inspiral, one should then use a lower value of $Q$, to better separate these two features in time. However, in doing so the frequency content becomes largely indistinguishable, and is can be difficult to discern that the time-frequency map even represents a gravitational wave signal. On the other hand, at higher values of $Q$, the frequency values are more strongly highlighted but the time localization becomes smeared such that the double-chirp pattern is indistinguishable from a single peak. An example of this effect is shown in Fig.(\ref{GT0568_morlet_Qvary}), where we display CWT scalograms with $Q = 1.0, 8.0, 16.0$. In testing a variety of options with the CWT over several example waveforms with this characteristic, values of $Q \in \left[4.0,6.0\right]$, corresponding to mother frequencies of $f_0 \in \left[0.45, 0.68\right]$ have been the most useful for analyzing the double chirp pattern. This range is consistent with the analysis in\cite{CalderonBustillo2020}, where a mother frequency of $f_0 = 0.4$ is used.\par

One can see that the scalogram in Fig.(\ref{GT0568_Q5}) reveals an undulating frequency structure during the inspiral, with a series of peaks that occur at higher and higher frequency as the system approaches merger. This is a consequence of the high mass ratio of the system; the asymmetry modulates the waveform, with each orbit effectively sweeping a burst of GW radiation across the observer's line of sight. One can view this frequency structure as a series of consecutive up-chirps and down-chirps, and considering the inner product perspective on CWT described in Sec.(\ref{CWT_section}), one is provoked to wonder what the time-frequency representation looks like using a different wavelet basis. To this end we can also use \emph{chirplets} in the CWT process: Morlet-Gabor wavelets that have their own frequency evolution. Further details on the chirplet basis are discussed in Appendix Sec.(\ref{appendix_chirplets}).\par

\begin{figure}
    \centering
    \includegraphics[width=3.5in]{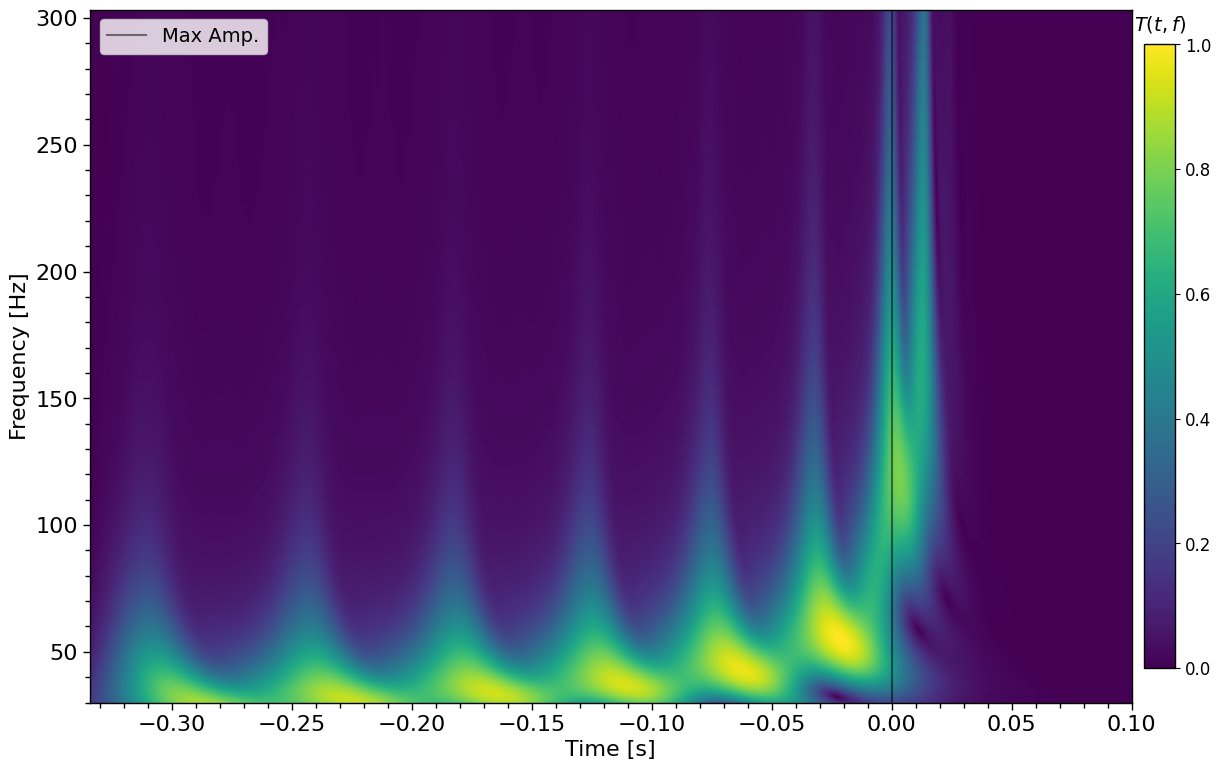}
    \caption{Frequency structure evolution in the noiseless, freespace gravitational waveform GT0568 $\left(q = 10.0,\: \abs{\mathbf{S}_1} = \abs{\mathbf{S}_2} = 0\right)$, scaled to $M = 80 \: M_{\odot}$ at $d_L = 500$ [Mpc], oriented at inclination $\iota = \pi / 2$ and phase $\phi = 0$, and generated from a starting frequency of 30.0 [Hz]. The black vertical line denotes the time location of the maximum waveform amplitude, which is used as an approximation for the time of merger. A chirplet basis with $d = 0.1$ has been used for the CWT analysis, with $Q = 5.0$.}
    \label{GT0568_chirplet}
\end{figure}

In Fig.(\ref{GT0568_chirplet}) one can see that in the chirplet basis, the CWT scalogram has several differences when compared to its MG wavelet counterpart in Fig.(\ref{GT0568_Q5}). As discussed in Appendix Sec.(\ref{appendix_chirplets}), the chirplet basis rotates burst-like signals in time-frequency space. Using a basis with an upward chirp rate (up-chirps) causes bursts to rotate counter-clockwise in time-frequency space, and using a basis with a downward chirp rate (down-chirps) causes bursts to rotate clockwise in time-frequency space. Additionally, for a basis of either up-chirps or down-chirps, bursts get stretched in frequency. Here we see that with an up-chirp CWT basis, each inspiral burst appears tilted counter-clockwise, and stretched in frequency. In a down-chirp CWT basis, each inspiral burst appears tilted clockwise, and is likewise stretched in frequency, although we omit the figure. The frequency-stretching effect can be mitigated to some degree by increasing $Q$ for the CWT analysis, but this does yield poorer time localization, and going too high on $Q$ can make the postmerger chirp feature indistinguishable from the merger peak.\par 

\begin{figure}
    \centering
    \includegraphics[width=3.5in]{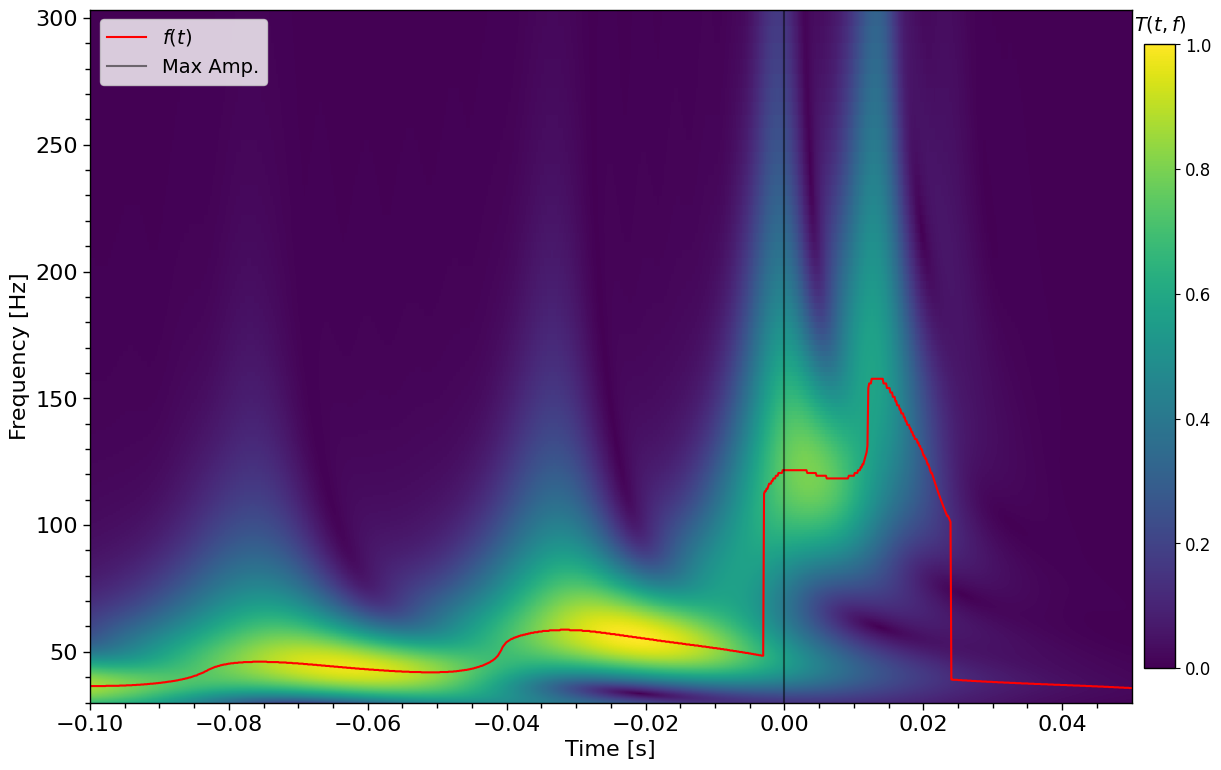}
    \caption{Frequency structure evolution in the postmerger of the noiseless, freespace gravitational waveform GT0568 $\left(q = 10.0,\: \abs{\mathbf{S}_1} = \abs{\mathbf{S}_2} = 0\right)$, scaled to $M = 80 \: M_{\odot}$ at $d_L = 500$ [Mpc], oriented at inclination $\iota = \pi / 2$ and phase $\phi = 0$, and generated from a starting frequency of 30.0 [Hz]. The black vertical line denotes the time location of the maximum waveform amplitude, which is used as an approximation for the time of merger. A chirplet basis with $d = 0.1$ has been used for the CWT analysis, with $Q = 6.0$. The red line traces the brightest pixel at each time step, yielding an effective one-dimensional $f(t)$.}
    \label{GT0568_chirplet_zoom}
\end{figure}

Shown in Fig.(\ref{GT0568_chirplet_zoom}) is a zoomed-in view of the postmerger of GT0568, rendered using CWT with a chirplet basis of $d = 0.1$ at $Q = 6.0$. Additionally, in this figure we trace an effective one-dimensional $f(t)$ by highlighting the brightest pixel in the time-frequency map (shown in red). It is for this reason that the chirplet basis is most useful when applied to this class of signal; one can see that there are clear boundaries that denote the inspiral peak, the postmerger peak, and the end of the frequency evolution. These boundaries allow one to clearly define the postmerger region of the waveform based on its time-frequency map, and also the existence of the postmerger frequency peak. Further discussion of this method, its use towards parameterizing the double-chirp pattern, and analysis of its physical meaning may be seen in our companion work~\cite{Henshaw2025}.\par

\subsection{BNS Postmerger}
\begin{figure}
    \centering
    \includegraphics[width=3.5in]{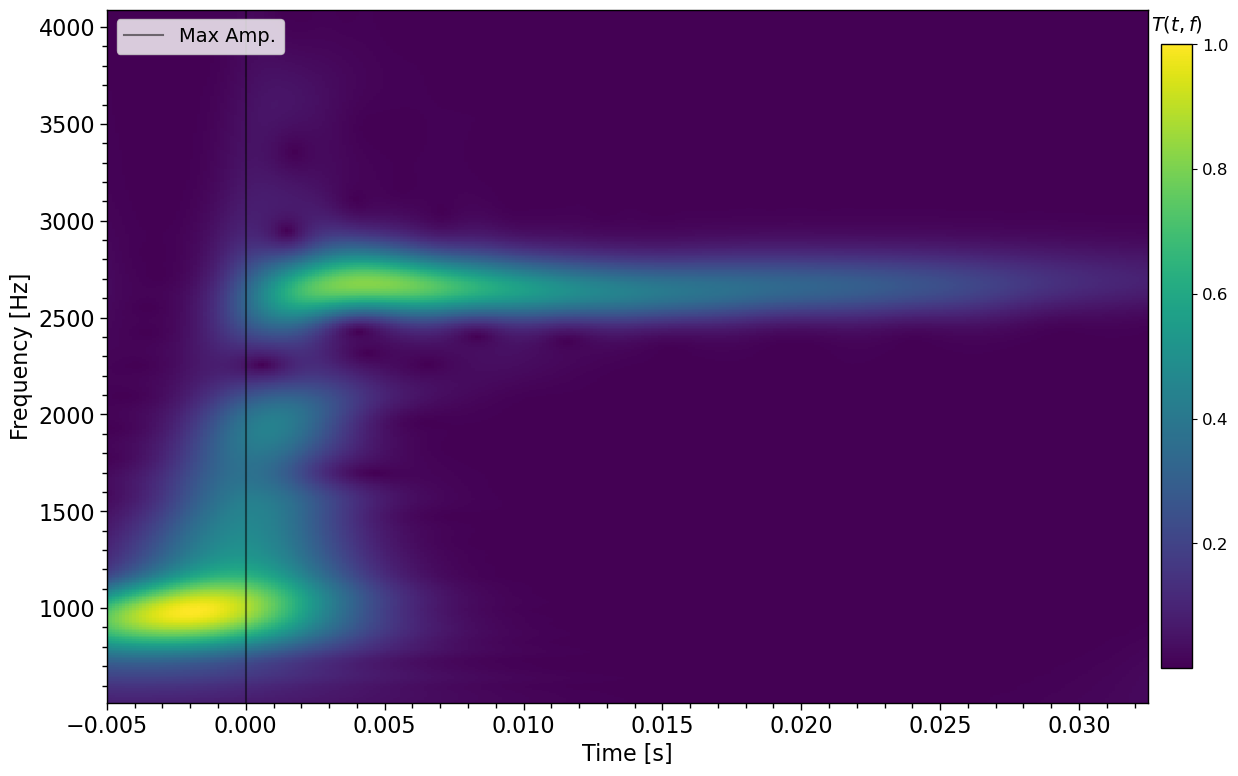}
    \caption{Frequency structure evolution in the noiseless, freespace BNS gravitational waveform, DD2 with component masses $m_1 = 1.35 M_\odot$ and $ m_2 = 1.35 M_\odot$ and a peak postmerger frequency $ f_{peak}\sim2500 [Hz]$. The black vertical line denotes the time location of the maximum waveform amplitude, which is used as an approximation for the time of merger.}
    \label{dd2_Q32}
\end{figure}

In the case of a BNS merger, oscillations of the postmerger remnant can emit a rich, high frequency gravitational wave signal. 
Given a postmerger remnant of a hypermassive or supramassive neutron star, the post-merger gravitational wave signal could be milliseconds long with high frequencies between 1-4 kHz \cite{Postmerger2, Chatziioannou:2017ixj}. 
With a confident detection, this postmerger signal could provide a new probe to study extremely dense nuclear matter given the the quasi-universal relations between spectral features of the signal and properties of the
neutron star \cite{EOSDEP, PMConstraints3Gen}. For example, \cite{EOSDEP, PhysRevLett.108.011101} found that there is a relationship between the peak frequency of the BNS postmerger signal and the radius of non-rotating NSs. This radius can then be used to help constrain the NS EoS through the one-to-one mapping from the mass-radius relationship of neutron stars to the EoS \cite{EOSDEP}.
Following a detection, one way to pinpoint the peak frequency of the postmerger signal is through a time-frequency map of the waveform. To visualize the BNS postmerger signal in the time-frequency plane, we use CWT. Shown in Fig.(\ref{dd2_Q32}) is the time-frequency map of the postmerger waveform for the DD2 EoS~\cite{2024arXiv240509513S} with $m_1 = 1.35 M_\odot$ and $ m_2 = 1.35 M_\odot$. One can see that following the initial frequency increase characteristic of a compact binary coalescence, there is excess power at approximately $2500 [Hz]$. See Sec.\ref{discussion} for additional discussion on the physical interpretation of this high frequency signature.\par

\begin{figure}
    \centering
    \includegraphics[width=3.5in]{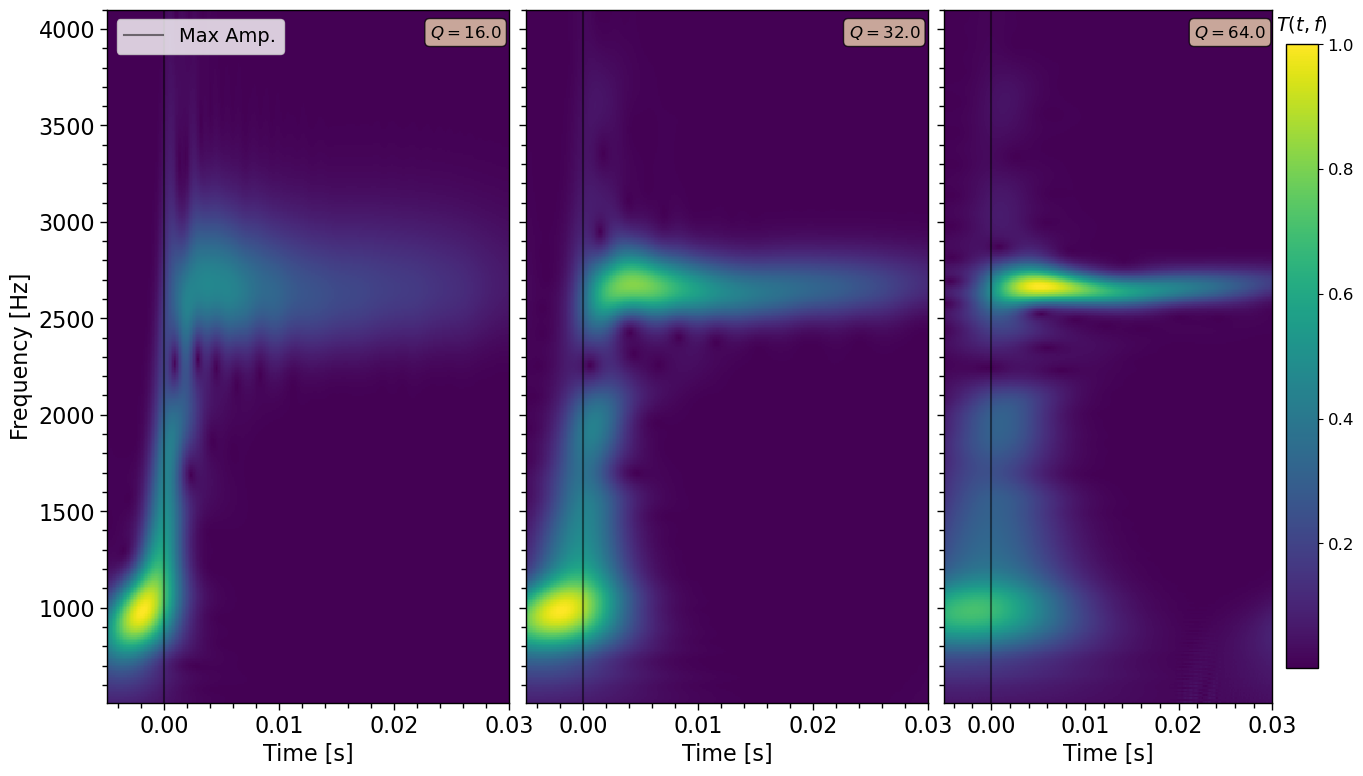}
    \caption{Quality factor variation for the CWT time-frequency map with the Morlet wavelet applied to the BNS waveform DD2. From left to right, $Q = 16.0, 32.0, 64.0$.}
    \label{dd2Qrange}
\end{figure}

The time-frequency map in Fig.(\ref{dd2_Q32}) was made with the Morlet wavelet at $Q=32.0$. As explained in Sec.\ref{CWT_section}, at higher $Q$ values, the frequency resolution is enhanced, but at the cost of diminished time localization. Given that peak frequency of a BNS postmerger signal could be used to help constrain the NS EoS, a higher value of $Q$ should be used to get a more confident measure of $f_{peak}$. In Fig.(\ref{dd2Qrange}), CWT was tested with varied $Q$ values of $Q \in \left[16.0,32.0,64.0\right]$, corresponding to mother frequencies of $f_0 \in \left[1.80,3.60,7.20\right]$. One can see that at lower $Q$ the postmerger oscillation is smeared in frequency, but as $Q$ increases it becomes more localized. However the localization of the merger is also smeared in time as $Q$ becomes large. For this reason $Q=32.0$ is preferred to preserve both characteristics.\par 

\section{Discussion}
\label{discussion}

The time-frequency maps above highlight the ability of the CWT to distinguish specific features in the frequency evolution of gravitational wave signals. In particular we have examined the postmerger waveforms from both BBH and BNS systems, where characteristic features are connected to the physics of their originating systems. The CWT method is highly flexible, and can be tuned to the specific problem at hand by varying chiefly the quality factor $Q$, and also the wavelet basis. The quality factor determines one's ability to localize features in either time or frequency, with simultaneous localization in both being limited by the Heisenberg-Gabor uncertainty relation. A lower value $Q$ will improve time localization at the cost of frequency localization, and vice versa for a higher value of $Q$. In the case of BBH systems with asymmetric mass ratio, where frequency features in the postmerger are evident when viewed from an edge-on inclination, a lower $Q$ is preferred. This allows one to differentiate the postmerger frequency peak from that of the inspiral; the exact frequency content of the postmerger peak is of secondary importance to its location in time.\par

The BNS postmerger case has the opposite requirements, as we are more interested in the frequency content than in the time localization. Here we therefore want a higher value of $Q$, to better isolate the frequency of the postmerger remnant oscillations. High frequency localization of the postmerger peak frequency would give way to constraints on the NS EoS through the empirical relationship between $f_{peak}$ and the radius of a non-rotating NS. As such, exact time localization is sacrificed for the purpose of determining the precise frequency content of the BNS postmerger signal. This work primarily examined noiseless, freespace gravitational waves, and demonstrated how CWT can be utilized to extract important frequency features embedded in the BNS postmerger signal. Given estimated merger rates, current and future detector sensitivities, and predicted waveform morphologies, the probability of near-future detection is low \cite{Clark2016, Torres-Rivas:2018svp, ModeStacking}. Given the low probability of detection, studies such as \cite{ModeStacking} and \cite{CriswellStacking} have explored combining observations using Bayesian statistics to increase the likelihood of detection. An additional stacking method that can be used to boost the detection sensitivity is to combine TF maps from multiple postmerger candidates created using CWT. This method will be detailed in a subsequent paper.\par

In addition to varying the quality factor, one can also use CWT with different bases of wavelets. Many applications in gravitational wave analysis use the Morlet-Gabor wavelet, as it is maximally compact in time-frequency space. Modifying the MG wavelet into a chirplet largely preserves this benefit, but offers an additional axis of exploration in the space of the signal's time-frequency representation. In the BBH case, where one is primarily interested in the separation between the inspiral and postmerger chirps, the chirplet basis allows one to clearly identify boundaries that denote the start and end of the postmerger stage of the signal. This enables the development of a consistent statistic that tracks the existence of the postmerger chirp over the angular variations of the observer orientation, which can then be used to predict the direction of maximal emission based on the intrinsic parameters of the waveform. This parameterization process and its application towards inferring the geometry of the final black hole horizon are detailed in our companion work~\cite{Henshaw2025}.\par

Beyond the specific applications to the BBH and BNS postmerger waveforms, there are other potential uses for the CWT framework. In this work, we chiefly considered freespace gravitational waves - simulated waveforms absent the realities of sensitivity and noise that must be considered for detectors like LIGO, Virgo, and KAGRA. In practice the multiresolution Q-scan~\cite{Chatterji2004, Chatterji2005}, which performs a CWT-like process over multiple quality factors, is very effective in distinguishing a gravitational wave signal (or other transient) from the background detector noise. However this method constitutes a more general approach, and lacks the specificity of time/frequency localization offered by the CWT framework. We speculate that a variation of the Q-scan algorithm and excess power method~\cite{Anderson2001} that incorporates the capabilities of CWT could be developed to search for specific features in localized time-frequency tiles. For now we leave this possibility to future work.\par

Additionally, the flexibility of CWT to incorporate further wavelets beyond the sine-Gaussian MG wavelets allows for the creation of time-frequency maps that have a fundamentally different structure. Herein we have explored the use of chirplets, which rotate structures in time-frequency space, for the BBH postmerger case. Chirplets, which add the chirp rate $d$ to the MG wavelet, therefore effectively add an extra axis to the space of possible time-frequency analyses on gravitational wave data. Included with this work is an example script for running CWT with chirplets on previous catalog events, which may be found at \href{https://github.com/chadhenshaw/gw_cwt}{\texttt{github.com/chadhenshaw/gw\_cwt}}; see Appendix Sec.(\ref{numerical_imp}) for further details. Additionally there are also many other wavelet bases that may be of interest; for example the ``Mexican hat" wavelet described in~\cite{Addison2002} may be useful for analysis of spurious noise transients in LVK data, as it bears similar time-frequency morphology to ``blip" glitches. In a similar vein, exponential shapelets~\cite{Berge2019} may be useful for transients that feature damped oscillations, or even for black hole ringdown. We leave the implementation and exploration of such methods and applications to future work.\par

\begin{center}
\noindent\rule{8cm}{0.4pt}    
\end{center}

\section*{Acknowledgements}

We thank Meg Millhouse and James Clark for helpful discussions and technical advice. This research has made use of data, software and/or web tools obtained from the Gravitational Wave Open Science Center (https://www.gw-openscience.org/), a service of LIGO Laboratory, the LIGO Scientific Collaboration and the Virgo Collaboration. LIGO Laboratory and Advanced LIGO are funded by the United States National Science Foundation (NSF) as well as the Science and Technology Facilities Council (STFC) of the United Kingdom, the Max-Planck-Society (MPS), and the State of Niedersachsen/Germany for support of the construction of Advanced LIGO and construction and operation of the GEO600 detector. Additional support for Advanced LIGO was provided by the Australian Research Council. Virgo is funded, through the European Gravitational Observatory (EGO), by the French Centre National de Recherche Scientifique (CNRS), the Italian Istituto Nazionale di Fisica Nucleare (INFN) and the Dutch Nikhef, with contributions by institutions from Belgium, Germany, Greece, Hungary, Ireland, Japan, Monaco, Poland, Portugal, Spain. KAGRA is supported by the Ministry of Education, Culture, Sports, Science, and Technology (MEXT) in Japan, and is hosted by the Institute for Cosmic Ray Research (ICRR), the University of Tokyo, and co-hosted by High Energy Accelerator Research Organization (KEK) and the National Astronomical Observatory of Japan (NAOJ). This material is based upon work supported by the LIGO Laboratory which is a major facility fully funded by the NSF. This work was supported by NSF grants PHY-1809572 \& PHY-2110481. This paper has LIGO document number P2400016.

\begin{center}
\noindent\rule{8cm}{0.4pt}    
\end{center}

\input{Appendix}

\centering
\noindent\rule{8cm}{0.4pt}

\bibliography{references}


\end{document}

%% file: Appendix.tex
\appendix

\setcounter{equation}{0}
\renewcommand{\theequation}{A\arabic{equation}}
\renewcommand{\thesection}{\ifnum\value{section}=1 A\else A\Roman{section}\fi}

\section{Appendix}
\label{appendix_main}
For a function $g(x)$, with power distribution $\abs{g(x)}^2$, its normalized $n$'th central moment is given by:

\begin{align}
    m_n = = \frac{\int_{-\infty}^{\infty} x^n \abs{g(x)}^2 dx}{\int_{-\infty}^{\infty} \abs{g(x)}^2 dx},
\end{align}

where $n=1$ corresponds to the mean value, and $n=2$ to the variance. If $g(x)$ is a time series $g(t)$ with corresponding Fourier transform $\hat{g}(f)$, its central time $\tau$, and a central frequency $\phi$ are given by the first central moments:

\begin{align}
    \tau &= \frac{\int_{-\infty}^{\infty} t \abs{g(t)}^2 dt}{\int_{-\infty}^{\infty} \abs{g(t)}^2 dt},\label{central_time}\\
    \phi &= \frac{\int_{-\infty}^{\infty} f \abs{\hat{g}(f)}^2 df}{\int_{-\infty}^{\infty} \abs{\hat{g}(f)}^2 df}.\label{central_freq}
\end{align}

The second central moments then give the characteristic duration $\sigma_t$ and bandwidth $\sigma_f$:

\begin{align}
\sigma_t^2 &= \frac{\int_{-\infty}^{\infty} \left(t - \tau\right)^2 \abs{g(t)}^2 dt}{\int_{-\infty}^{\infty} \abs{g(t)}^2 dt}\label{characteristic_duration}\\
\sigma_f^2 &= \frac{\int_{-\infty}^{\infty} \left(f - \phi\right)^2 \abs{\tilde{g}(f)}^2 df}{\int_{-\infty}^{\infty} \abs{\tilde{g}(f)}^2 df}\label{bandwidth}
\end{align}

\section{Wavelets}
\label{appendix_wavelets}

A wavelet is a localized wave-like function $\psi(t)$ that satisfies three criteria. First, a wavelet must have finite energy $E$:

\begin{align}
    E = \int_{-\infty}^{\infty} \abs{\psi(t)}^2 dt \: < \: \infty.
\end{align}

Note that here the vertical brackets represent the modulus operator - i.e if $\psi(t)$ is a complex function, then $\abs{\psi(t)}^2 = \psi^*(t)\psi(t)$, where $\psi^*(t)$ is the complex conjugate of $\psi(t)$. In practice, it is common for the wavelet to further be normalized such that it has unit energy. Secondly, given the Fourier transform of the wavelet $\hat{\psi}(f)$, the wavelet must have no zero-frequency component:

\begin{align}
    C_g = \int_{0}^{\infty} \frac{\abs{\hat{\psi}(f)}^2}{f} df  \: < \: \infty.
\end{align}

This criterion is referred to as the \emph{admissibility condition}, and $C_g$ is called the \emph{admissibility constant}, whose value depends on the specific wavelet. The final criterion applies to complex wavelets - their Fourier transform must be real, and must be zero for negative frequencies. Any wave-like function that satisfies these three criteria is a wavelet, and may be used in the continuous wavelet transform. For gravitational wave analysis, the most common such function is the Morlet wavelet:

\begin{align}
    \psi(t, f_0) &= \pi^{-\frac{1}{4}} \exp\left[-\frac{t^2}{2}\right]\left(\exp\left[i 2 \pi f_0 t \right] - \exp\left[- \left(2 \pi f_0\right)^2/2\right]\right).
    \label{appendix_morlet_withcorrection}
\end{align}

Here the second exponential term within the parentheses is referred to as the correction term, as it ensures the wavelet satisfies the second criterion by giving it zero mean. However in practice this term is discarded, as it becomes negligible for $f_0 \gg 0$\footnote{In the numerical implementation of CWT, we set a lower frequency cutoff of $f_{eff} \geq 0.4$ for this reason.}. We will still refer to the resultant approximation as a wavelet, as it only violates the admissibility condition for low frequencies. This approximation is often referred to as the Gabor wavelet, or the Morlet-Gabor (MG) wavelet:

\begin{align}
    \psi(t, f_0) &= \pi^{-\frac{1}{4}} \exp\left[-\frac{t^2}{2}\right]\exp\left[i 2 \pi f_0 t \right] \label{appendix_MGwavelet_timedomain}\\
    \tilde{\psi}(f, f_0) &= \pi^{\frac{1}{4}} \sqrt{2} \exp\left[-\frac{1}{2} \left(2\pi f - 2\pi f_0\right)^2\right]. \label{appendix_MGwavelet_freqdomain}    
\end{align}

Note that the MG wavelet has three components: an amplitude $\pi^{-\frac{1}{4}}$, a Gaussian window $\exp\left[-\frac{t^2}{2}\right]$ with unit standard deviation, and a complex sinusoid $\exp\left[i 2 \pi f_0 t \right]$ with frequency $f_0$. Its central time is identically zero, and its central frequency is $f_0$. Its characteristic duration is $\sigma_t = 1 / \sqrt{2}$, and its bandwidth is $1 / 2\sqrt{2} \pi$. The duration-bandwidth product of the MG wavelet is thus $\sigma_t \sigma_f = 1 / 4\pi$, which is the minimum value in the Heisenberg-Gabor uncertainty principle. In this way, MG wavelets are maximally compact in time-frequency space, which makes them ideal for time-frequency analysis methods like the continuous wavelet transform (Eq.(\ref{CWT})). In this method, the wavelet's scale is adjusted by the dilation parameter $a$, and it's time localization is shifted by the translation parameter $b$:

\begin{align}
    \psi(t; f_0, a, b) &= \pi^{-\frac{1}{4}} a^{-1/2} \exp\left[-\frac{1}{2}\left(\frac{t-b}{a}\right)^2\right]\cdot\nonumber\\
    &\cdot\exp\left[i 2 \pi f_0 \left(\frac{t-b}{a}\right) \right].
    \label{appendix_MGwavelet_diltran}
\end{align}

The wavelet's energy is normalized at every scale:

\begin{align}
    \int_{-\infty}^{\infty} \abs{\psi(t)}^2 dt &= \frac{1}{a\sqrt{\pi}} \int_{-\infty}^{\infty} \exp\left[-\left(\frac{t-b}{a}\right)^2\right] dt = 1.
\end{align}

This normalization is maintained in the frequency domain; first we compute the Fourier transform of the dilated and translated MG wavelet:

\begin{align}
    \hat{\psi}(f) &= \int_{-\infty}^{\infty} \psi(t) \exp\left[-2 \pi i f t\right] dt\nonumber\\
    &= \sqrt{2} \pi^{\frac{1}{4}} a^{1/2} \exp\left[-i 2\pi f b\right]\exp\left[-\frac{1}{2}\left(2\pi f_0 - 2\pi a f\right)^2\right],
\end{align}

Note that in the untranslated, non-dilated case $\left(b=0, a=1\right)$, one recovers exactly the Fourier transform of the MG wavelet in Eq.(\ref{appendix_MGwavelet_freqdomain}). The energy is then:

\begin{align}
    \int_{-\infty}^{\infty} \abs{\tilde{\psi}(f)}^2 df = 1,
\end{align}

The denominators in Eqs.(\crefrange{central_time}{bandwidth}) for the MG wavelet are thus unity, and we can proceed to calculate the central time:

\begin{align}
    \tau &= \int_{-\infty}^{\infty} t \abs{\psi(t)}^2 dt = b,
\end{align}

which returns as the translation parameter $b$, as expected. Now we calculate the central frequency:

\begin{align}
    \phi &= \int_{-\infty}^{\infty} f \abs{\tilde{\psi}(f)}^2 df = \frac{f_0}{a}.
\end{align}

Note that in the case of unit scale, i.e. $a=1$, we recover the central frequency as exactly $f_0$, as one expects. Next we will calculate the characteristic duration:

\begin{align}
    \sigma_t^2 &= \int_{-\infty}^{\infty} \left(t - \tau\right)^2 \abs{\psi (t)}^2 dt = \frac{a}{\sqrt{2}},
\end{align}

which gives an indication of how localized the wavelet is in the time domain. We see that this is directly proportional to the dilation parameter; thus at larger scales the wavelet is more spread out in time. The bandwidth is then:

\begin{align}
    \sigma_f^2 &= \int_{-\infty}^{\infty} \left(f - \phi\right)^2 \abs{\tilde{\psi}(f)}^2 df = \frac{1}{2\sqrt{2} \pi a}.
\end{align}

We obtain a bandwidth - and thus a frequency resolution - that is inversely proportional to the wavelet scale. Therefore at larger scales the wavelet has a narrower bandwidth, as one would expect - more frequencies are encapsulated by the larger wavelet. The quality factor $Q$ is then defined as the ratio of central frequency to bandwidth:

\begin{align}
    Q \equiv \frac{\phi}{\sigma_f} = 2\sqrt{2} \pi f_0,
\end{align}

which we see does not depend on the wavelet scale; $Q$ depends only on the starting frequency of the sinusoid. We also see that the duration-bandwidth product is still $\sigma_t \sigma_f = 1 / 4\pi$, so dilated MG wavelets remain maximally compact in time-frequency space. Now considering that the quality factor does not depend on wavelet scale, we can rewrite the equation for the dilated and translated Morlet wavelet as:

\begin{align}
    &\psi(t; Q, a, b) =\nonumber\\
    &= \pi^{-\frac{1}{4}} a^{-1/2} \exp\left[-\frac{1}{2}\left(\frac{t-b}{a}\right)^2\right]\exp\left[i \frac{Q}{\sqrt{2}} \left(\frac{t-b}{a}\right) \right],
\end{align}

where we see that for a given dilation factor $a$ we have the effective sinusoid frequency:

\begin{align}
    f_{eff} = \frac{Q}{2\sqrt{2} \pi a} = \frac{f_0}{a},
\end{align}

which is equivalent to the central frequency $\phi$. When running CWT, it is this effective frequency that you are interrogating when convolving the wavelet with the signal. 

\section{Chirplets}
\label{appendix_chirplets}

The central frequency of the MG wavelet in Sec.(\ref{appendix_wavelets}) above is stationary, but may be modified to evolve in time by including a chirp rate parameter $d$ (following the notation of \cite{Mohapatra2012}) such that the central sinusoid frequency $f_0$ becomes:

\begin{align}
    f_0(t) = f_0 + d*t,
\end{align}

and the mother wavelet, now a \emph{chirplet}, becomes:

\begin{align}
    \psi(t; f_0, d, a, b) &= \pi^{-\frac{1}{4}} \exp\left[-\frac{1}{2}\left(\frac{t-b}{a}\right)^2\right]*\nonumber\\
    &*\exp\left[i 2 \pi \left(f_0 \left(\frac{t-b}{a}\right) + \frac{d}{2} \left(\frac{t-b}{a}\right)^2 \right) \right],\nonumber\\
    \psi(t; Q, a, b) &= \pi^{-\frac{1}{4}} \exp\left[-\frac{1}{2}\left(\frac{t-b}{a}\right)^2\right]*\nonumber\\&*\exp\left[i\left( \frac{Q}{\sqrt{2}} \left(\frac{t-b}{a}\right) + \pi d \left(\frac{t-b}{a}\right)^2 \right) \right].
\end{align}

Note that here the chirp parameter $d$ is a fixed quantity, and as such in the CWT process becomes another parametric input for the mother wavelet. A positive value of $d > 0$ corresponds to an \emph{up-chirp}, where the wavelet frequency increases over time, and a negative value of $d > 0$ corresponds to a \emph{down-chirp}, where the wavelet frequency decreases over time. Note that if $d=0$ we recover exactly the Morlet-Gabor wavelet. Also note that this parameterization of the chirp rate is different than that in e.g.\cite{Millhouse2018}, where the chirp parameter $\beta = 2d$ is used. Below we create a variation of Fig.2 from that work, using the CWT process to create the time-frequency map of different chirplets.

\begin{figure}
    \centering
    \includegraphics[width=3.5in]{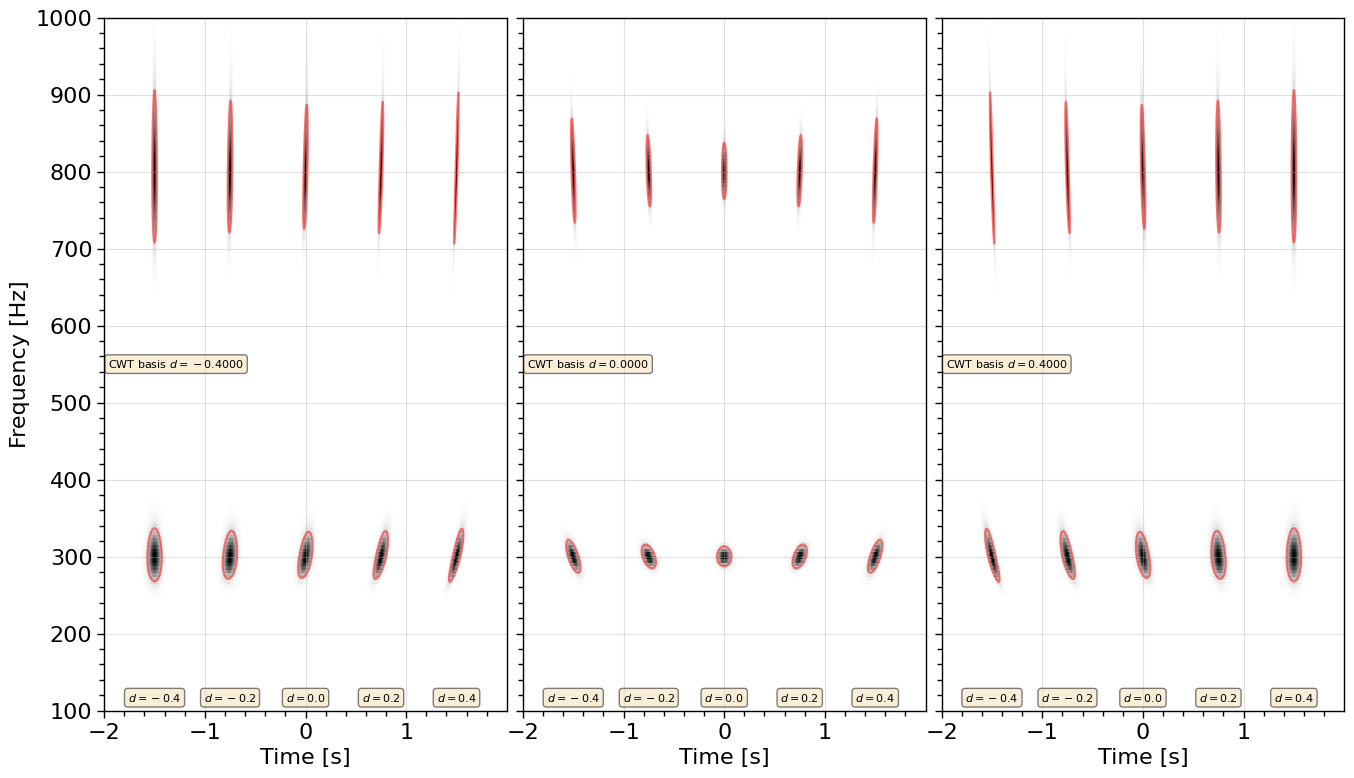}
    \caption{Time-frequency representation of different chirplets, created using different chirplet bases with CWT. From left to right, the CWT process was used with a chirp rate of $d = \left[-0.4, 0.0, 0.4\right]$, with $Q=64$ in each case. Each panel contains ten different chirplet signals. The bottom row in each panel contains chirplets at $f_0 = 300$ [Hz] and $Q=100$, and the top row contains chirplets at $f_0 = 800$ [Hz] and $Q=80$. From left to right, the chirplets in each panel have chirp rates of $d = \left[-0.4, -0.2, 0.0, 0.2, 0.4\right]$.}
    \label{chirplet_multi-map}
\end{figure}

Fig.(\ref{chirplet_multi-map}) was constructed by creating a series of ten chirplet signals with different parameters, running CWT on each signal, then displaying together all ten time-frequency maps. Each chirplet can be thought of as a short burst-like signal. The middle panel shows the representation of the ten chirplets using the MG wavelet (i.e. $d = 0.0$) basis with CWT. Here one can see that the time-frequency representation of each burst is an ellipse, as as the chirp rate becomes more positive (negative), the ellipse is rotated clockwise (counter-clockwise). The left and right panels show the same analysis, but with a chirplet basis at $d=-04$ and $d=0.4$ respectively. One can see that in this basis, the corresponding chirplet signal is counter-rotated in time-frequency space.\par

Consider first the right panel in Fig.(\ref{chirplet_multi-map}). The right-most chirplet signal $\left(d=0.4\right)$, which is tilted clockwise in the middle panel, has been rotated counter-clockwise to a vertical position. This is because the underlying chirp rate of the CWT basis matches the chirp rate of the signal, and as such its frequency structure has been localized in time. The remaining bursts, which all have a chirp rate less than that of the CWT basis, have also been rotated counter-clockwise, and their frequency localization has been stretched in accordance with the wider bandwidth of the CWT basis. The same principle applies to the left panel in Fig.(\ref{chirplet_multi-map}), but with the opposite sense. The bursts with chirp rates greater than that of the underlying CWT basis are now rotated clockwise, countering their natural rotation in time-frequency space.\par

\section{Numerical implementation of CWT}
\label{numerical_imp}

To create the time-frequency maps herein, we use a modified version of a previous Python implementation of CWT~\cite{pyCWT}. The modified code, which we refer to as \texttt{gw\_cwt}, is available at \href{https://github.com/chadhenshaw/gw_cwt}{\texttt{github.com/chadhenshaw/gw\_cwt}} along with example scripts. The primary function in this package is \texttt{build\_cwt}, which accepts as input a timeseries array and corresponding uniformly-sampled timestamps, runs the \texttt{cwt} function, and returns a dictionary containing properties of the resultant time-frequency map. The \texttt{cwt} function, which is a time domain convolution, is implemented as a product between signal and wavelet in the frequency domain. At each wavelet scale, the timeseries and wavelet are converted to the frequency domain by FFT, and the inverse FFT of their product is then computed to complete the convolution.\par

The \texttt{build\_cwt} function also contains a number of options that the user may specify to customize their analysis, with the primary variables being the quality factor $Q$, and the chirp rate $d$. The number of frequencies interrogated is determined by the option \texttt{n\_conv}, which is the number of convolutions. Increasing this setting samples more frequencies within the given range at the cost of compute time. Given a range of frequencies, wavelet scales are computed as $a = f_0 f_s / f_{eff}$, where $f_s$ is the sampling rate of the timeseries. Additionally, the Nyquist frequency $f_s / 2$ determines the upper bound on the signal frequency one can interrogate. By default, the program will create a linearly-spaced array of wavelet scales, which corresponds to a logarithmically-spaced array of frequencies. The user may instead request a linearly-spaced frequency array, in which case the program will create the corresponding logarithmically-spaced array of wavelet scales.  Alternatively for each case, the user may specify the spacing between frequencies or scales as $df$ or $da$, which supersedes the number of convolutions. Finally, the user may instead input directly an array of frequencies to compute over.\par

Included also in the \texttt{gw\_cwt} package is the script \texttt{cwt\_catalog.py}, which simplifies the process of running CWT on GW data. Within this program, the \texttt{get\_data} function pulls gravitational wave data from the Python-based \texttt{pycbc.catalog}~\cite{alex_nitz_2023_10137381}. The data are downloaded and stored in a local \texttt{hdf5} compressed file, the organizational structure of which resembles a Python dictionary. The \texttt{run\_cwt} function then computes the time-frequency map for the downloaded event data, given user input of CWT parameters, and appends the results to the \texttt{hdf5} file. Finally, \texttt{plot\_cwt} is a function for plotting the time-frequency map. Examples and further documentation are available at \href{https://github.com/chadhenshaw/gw_cwt}{\texttt{github.com/chadhenshaw/gw\_cwt}}.\par

